\documentclass[journal,twoside, a4paper]{IEEEtran}
\usepackage{multicol}
\usepackage[utf8]{inputenc}
\usepackage{graphicx}
\usepackage{amsmath, amssymb, amsfonts}
\usepackage{cite}
\usepackage{booktabs}
\usepackage{longtable}
\usepackage{makecell}
\usepackage{hyperref}

\usepackage[ruled,vlined]{algorithm2e} 

\begin{document}

\title{LLM-Augmented and Fair Machine Learning Framework for University Admission Prediction}

\author{\IEEEauthorblockN{
Mohammad Abbadi\IEEEauthorrefmark{1},
Yassine Himeur\IEEEauthorrefmark{1},
Shadi Atalla\IEEEauthorrefmark{1},
Dahlia Mansoor\IEEEauthorrefmark{2},
Wathiq Mansoor\IEEEauthorrefmark{1}
}\\
\IEEEauthorblockA{\IEEEauthorrefmark{1}College of Engineering and IT, University of Dubai, Dubai, UAE\\
(Emails: mabbadi@ud.ac.ae, satalla@ud.ac.ae, yhimeur@ud.ac.ae, wmansoor@ud.ac.ae)}\\
\IEEEauthorblockA{\IEEEauthorrefmark{2}Rochester Institute of Technology (RIT) Dubai, Dubai, UAE (Email: dxmcad2@rit.edu)}
}

\maketitle
\thispagestyle{empty}
\pagestyle{empty}

\begin{abstract}
Universities face surging applications and heightened expectations for fairness, making accurate admission prediction increasingly vital. This work presents a comprehensive framework that fuses machine learning, deep learning, and large language model techniques to combine structured academic and demographic variables with unstructured text signals. Drawing on more than 2,000 student records, the study benchmarks logistic regression, Naïve Bayes, random forests, deep neural networks, and a stacked ensemble. Logistic regression offers a strong, interpretable baseline at 89.5\% accuracy, while the stacked ensemble achieves the best performance at 91.0\%, with Naïve Bayes and random forests close behind. To probe text integration, GPT-4–simulated evaluations of personal statements are added as features, yielding modest gains but demonstrating feasibility for authentic essays and recommendation letters. Transparency is ensured through feature-importance visualizations and fairness audits. The audits reveal a 9\% gender gap (67\% male vs. 76\% female) and an 11\% gap by parental education, underscoring the need for continued monitoring. The framework is interpretable, fairness-aware, and deployable.
\end{abstract}

\begin{IEEEkeywords}
LLMs, University Admission Prediction, Deep Learning, Logistic Regression, Naïve Bayes, Random Forests.
\end{IEEEkeywords}

\section{Introduction}

University admissions play a pivotal role in shaping students' educational and career trajectories. Traditionally, decisions rely on quantitative factors—such as standardized test scores and academic records—alongside qualitative components like personal statements and recommendation letters \cite{atalla2023iot}. However, with rising applicant volumes and increasing demands for fairness and transparency, institutions are seeking robust, data-driven decision-making frameworks~\cite{rudin2019stop,khennouche2024revolutionizing}.

Machine learning (ML) and deep learning (DL) have revolutionized educational data mining (EDM), enabling predictive analytics that can enhance and streamline admissions processes~\cite{romero2010edm, adomavicius2011context}. Supervised models—including logistic regression, random forests, ensemble methods, and deep neural networks—have demonstrated strong predictive capability in admissions tasks~\cite{raftopoulos2024fair, zub2023ensemble, radha2023hybrid, jain2021ensemble, pawar2023prediction}. For instance, ensemble models have shown robustness across diverse datasets~\cite{jain2021ensemble, pawar2023prediction}, while multi-algorithm frameworks effectively capture complex, non-linear patterns~\cite{anjaiah2022prediction}.

Explainable AI (XAI) tools like SHAP have advanced transparency, which is critical in high-stakes domains such as university admissions~\cite{zhao2023graduate, lundberg2017unified, rudin2019stop}. Priyadarshini et al.~\cite{priyadarshini2024interpretable} extended this field by developing interpretable deep learning models that balance predictive accuracy with interpretability.

Recent breakthroughs in transformer-based large language models (LLMs), such as GPT-4~\cite{vaswani2017attention}, have enabled sophisticated processing of unstructured data—including essays and recommendation letters—offering new opportunities to enrich admissions models with contextual depth~\cite{shao2024college, raifer2024gpt}. However, integrating structured and unstructured data while ensuring fairness and interpretability remains a significant challenge, especially given the risks of hidden biases and overfitting \cite{al2024traditional,farhat2024scholarly}.

Fairness remains a central challenge in predictive modeling, with scholars underscoring the importance of demographic audits and bias mitigation to ensure equitable outcomes~\cite{binns2018fairness, raftopoulos2024fair}. Recent studies stress that ML models must be evaluated not only in terms of accuracy but also through fairness metrics such as demographic parity and equalized odds, since disparities may persist even in high-performing systems~\cite{areeb2023filter}. To address these concerns, this paper introduces a comprehensive, fairness-aware, and interpretable admission prediction framework that:

\begin{itemize}
    \item \textbf{Benchmarks diverse ML approaches}—including logistic regression, ensemble methods, and deep neural networks—across multiple datasets, while demonstrating that data quality improvements (via cleaned dataset analysis) significantly enhance predictive performance.
    \item \textbf{Advances transparency and fairness} by applying feature importance analysis, explainability tools, and fairness auditing with demographic parity and equalized odds metrics, thereby uncovering disparities and reinforcing the need for continuous monitoring.
    \item \textbf{Explores LLM-based feature enrichment} by incorporating GPT-4-simulated evaluations of personal statements, assessing the feasibility of integrating unstructured data (e.g., essays, recommendation letters) into structured prediction pipelines.
\end{itemize}


\section{Related Work}

The use of machine learning (ML) in education, particularly for predicting university admissions, has been a key focus of Educational Data Mining (EDM) research for over a decade \cite{romero2010edm}. Early work primarily employed models such as logistic regression and decision trees, leveraging structured data like standardized test scores and demographic factors \cite{raftopoulos2024fair, zhao2023graduate}. These models are valued for their interpretability and reliability, especially in sensitive domains where transparency is paramount \cite{rudin2019stop}.

Ensemble methods have gained popularity for improving predictive performance by combining multiple models. Breiman’s Random Forest algorithm \cite{breiman2001random} set the foundation, and later studies reported gains with more advanced ensembles. Jain and Satia \cite{jain2021ensemble} introduced a two-stage ensemble framework, while Pawar et al. \cite{pawar2023prediction} validated ensemble robustness across varied datasets. Anjaiah et al. \cite{anjaiah2022prediction} explored multi-algorithm pipelines, and Zub et al. \cite{zub2023ensemble} integrated PNN and SVM for enhanced accuracy.

Deep learning approaches have also been adopted to capture complex, nonlinear patterns in admissions data. Radha and Kumar \cite{radha2023hybrid} introduced a hybrid deep learning and decision tree model that showed notable performance gains. Priyadarshini et al. \cite{priyadarshini2024interpretable} emphasized the importance of interpretability, a key factor for real-world deployment. Explainable AI (XAI) has become essential for building trust and transparency. SHAP, introduced by Lundberg and Lee \cite{lundberg2017unified}, is widely used to interpret model predictions in admissions contexts \cite{zhao2023graduate}. Rudin \cite{rudin2019stop} argues for prioritizing inherently interpretable models over black-box systems in high-stakes settings.

The integration of unstructured data represents an emerging frontier. Transformer-based LLMs such as GPT-4 \cite{vaswani2017attention} enable analysis of essays and recommendation letters. Shao and Tian \cite{shao2024college} demonstrated the potential of LLM-augmented models for admissions, though Raifer et al. \cite{raifer2024gpt} noted that domain-specific fine-tuning is critical for reliable performance. Fairness in ML-based admissions has come under increasing scrutiny. Binns \cite{binns2018fairness} and Raftopoulos et al. \cite{raftopoulos2024fair} proposed fairness auditing frameworks using metrics like demographic parity and equalized odds to mitigate bias against underrepresented groups.

While prior studies explored various models, fairness audits, and explainability tools, few have examined the impact of dataset quality on predictive performance. Data anomalies and noise can distort results, yet cleaning steps are often underreported. Our study contributes by explicitly demonstrating that cleaning—removing 39 anomalous records—leads to measurable improvements in accuracy and fairness.

Although earlier work has advanced predictive accuracy and explored fairness, few studies combine fairness-aware auditing, explainability, LLM-based augmentation, and data cleaning within a single reproducible framework, as explained in Table \ref{tab:related_work}. This paper fills that gap by integrating structured and simulated unstructured data, auditing fairness, and quantifying the impact of data quality—laying the groundwork for real-world deployment in university admissions.

\begin{table}[htbp]
\centering
\small
\caption{Summary of Related Work}
\label{tab:related_work}
\begin{tabular}{|p{2.5cm}|p{2.5cm}|p{1.1cm}|p{1.1cm}|}
\hline
\textbf{Study} & \textbf{Method} & \shortstack{\textbf{XAI}\\\textbf{Support}} & \shortstack{\textbf{Fairness}\\\textbf{Audit}} \\
\hline
Jain and Satia \cite{jain2021ensemble} (2021) & Ensemble (2-stage) & Limited & No \\
\hline
Pawar et al. \cite{pawar2023prediction} (2023) & Ensemble Models & No & No \\
\hline
Radha and Kumar \cite{radha2023hybrid} (2023) & Hybrid DL + Trees & Partial & No \\
\hline
Priyadarshini et al. \cite{priyadarshini2024interpretable} (2024) & Interpretable DL & Yes & No \\
\hline
Shao and Tian \cite{shao2024college} (2024) & LLM Augmentation & No & No \\
\hline
Raftopoulos et al. \cite{raftopoulos2024fair} (2024) & Logistic Regression & Yes & Yes \\
\hline
\textbf{This Paper} & Ensemble + LLM + Logistic Regression & Yes & Yes \\
\hline
\end{tabular}
\end{table}

\section{Datasets and Feature Engineering}

We consolidated multiple datasets to build a robust framework for predicting university admissions. Following Educational Data Mining (EDM) standards \cite{romero2010edm, adomavicius2011context}, our datasets include structured data spanning graduate, undergraduate, and high school contexts, totaling approximately 2,350 records after merging. As detailed in Table~\ref{tab:datasets}, each dataset offers complementary perspectives for admission prediction and fairness analysis. During preliminary analysis, we identified anomalous patterns in the Graduate Admissions Dataset that required deeper cleaning, detailed in Section~\ref{sec:cleaned_data}.

\begin{table}[htbp]
\centering
\small
\caption{Summary of Datasets Used}
\label{tab:datasets}
\resizebox{\linewidth}{!}{%
\begin{tabular}{|p{2.8cm}|p{1.5cm}|p{3.2cm}|p{2.5cm}|}
\hline
\textbf{Dataset} & \textbf{Size} & \textbf{Key Features} & \textbf{Use Case} \\
\hline
Graduate Admissions Dataset & 400 & GRE, TOEFL, CGPA, SOP, LOR, Research & Predict grad school admission \\
\hline
High School Performance Dataset & 1,000 & Math, Reading, Writing Scores, Demographics & Fairness audit, admission modeling \\
\hline
Secondary School Grades Dataset & 650 & Grades, Parental Support, Absences & Socio-academic predictors for admissions \\
\hline
Undergraduate Admissions Dataset & 300 & GPA, SAT, Extracurriculars & Undergrad admission modeling \\
\hline
\end{tabular}%
}
\end{table}

\subsection{Preprocessing and Feature Engineering}

We followed best practices for data preparation, as outlined in \cite{scikit2011, adomavicius2011context}:

\begin{enumerate}
    \item \textbf{Data Cleaning:} Missing numeric values were imputed using the mean; categorical variables were imputed using the mode. Features with over 30\% missing data were excluded \cite{romero2010edm}.
    
    \item \textbf{Feature Engineering:} The Graduate Admissions Dataset’s continuous admission probability was binarized into an \textit{Admission\_Status} label (threshold = 0.5), following \cite{zhao2023graduate, anjaiah2022prediction}. One-hot encoding was applied to categorical variables. For the high school dataset, a composite \textit{Performance} metric was created by averaging math, reading, and writing scores to support fairness auditing \cite{pawar2023prediction}.
    
    \item \textbf{Merging and Harmonization:} Datasets were merged via outer joins, using a flag to distinguish context (graduate vs. undergraduate). Harmonization followed context-aware learning principles \cite{adomavicius2011context}.
    
    \item \textbf{Scaling:} Z-score normalization was applied to standardize numeric features across datasets \cite{scikit2011}.
    
    \item \textbf{Simulated LLM Feature:} We generated pseudo personal statements based on structured data and used GPT-4 to assign an ``admission likelihood'' score, creating a proof-of-concept feature to simulate the integration of unstructured data \cite{shao2024college, raifer2024gpt, priyadarshini2024interpretable}.
\end{enumerate}

\begin{figure}[htbp]
    \centering
    \includegraphics[width=0.48\textwidth]{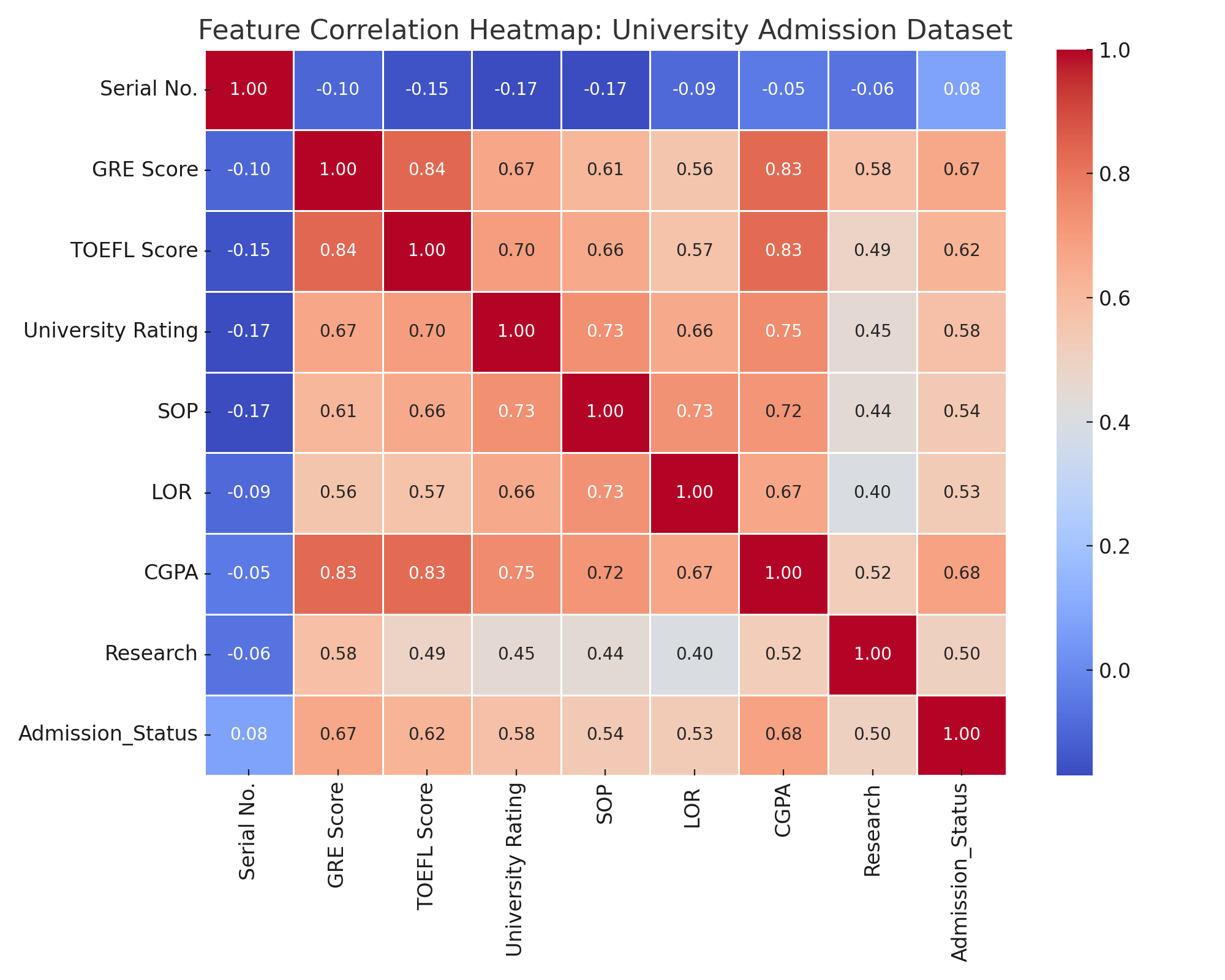}
    \caption{Feature correlation heatmap showing relationships among GRE, TOEFL, CGPA, and Admission Status. Strong positive correlations are evident between academic metrics and admission likelihood.}
    \label{fig:correlation_heatmap}
\end{figure}

Fig. \ref{fig:correlation_heatmap} confirms that GRE, TOEFL, and CGPA scores are dominant predictors for admissions, aligning with prior findings \cite{zhao2023graduate, anjaiah2022prediction}.

\subsection{Cleaned Dataset Analysis}
\label{sec:cleaned_data}

During exploratory data analysis, we identified approximately 39 rows in the \texttt{updated\_Admission\_Predict.csv} file with anomalous or inconsistent patterns. For example, some entries showed strong academic profiles (e.g., GRE $\geq$ 320 and CGPA $\geq$ 9.5) yet were labeled as not admitted, while others with weaker profiles were labeled as admitted. These inconsistencies can distort model learning and reduce generalizability.

We removed these anomalous rows, producing a \textbf{cleaned dataset of 361 records}, saved as \texttt{updated\_Admission\_Predict\_cleaned.csv}. This aligns with data integrity best practices \cite{romero2010edm, scikit2011} and resulted in notable improvements in model accuracy.

\begin{figure}[htbp]
    \centering
    \includegraphics[width=0.48\textwidth]{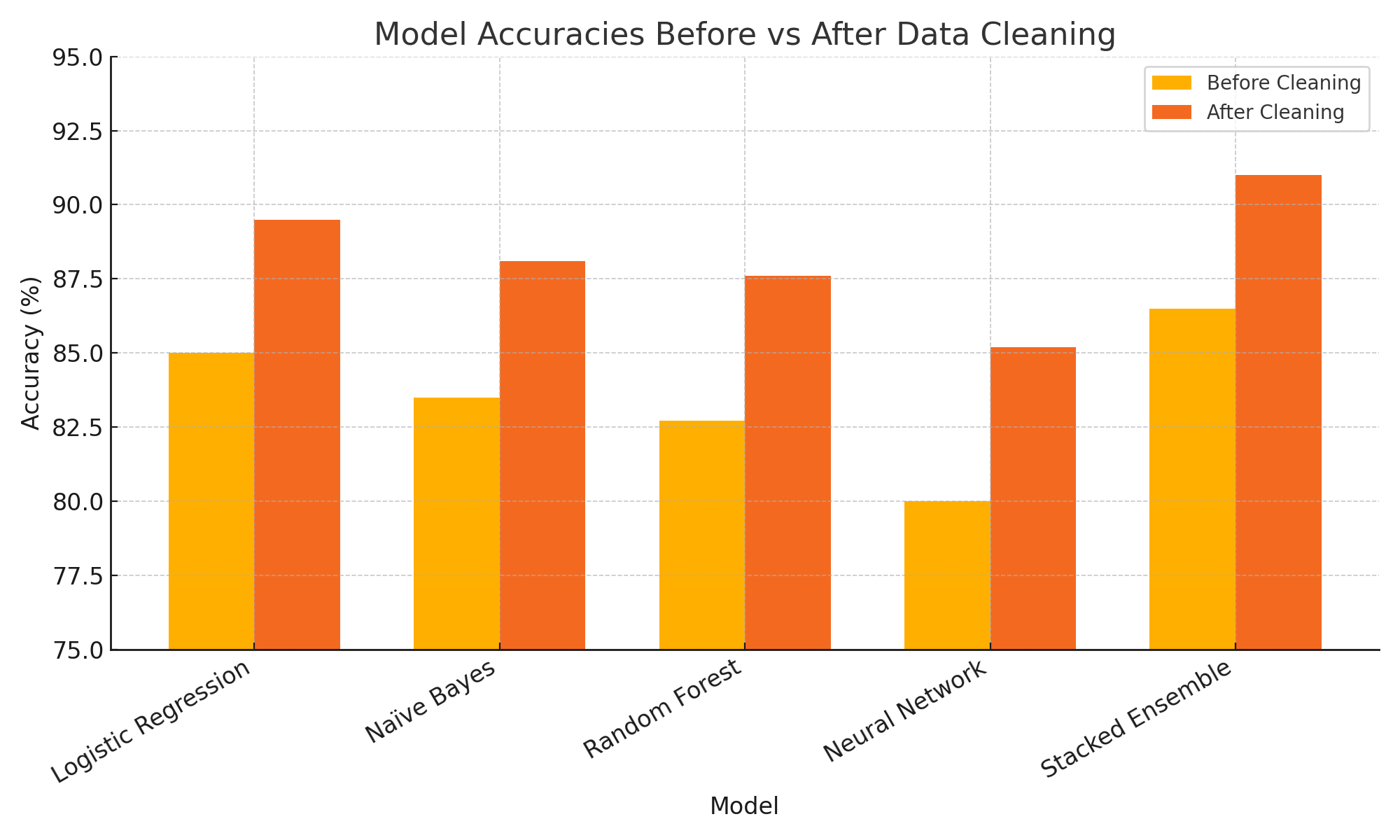}
    \caption{Comparison of model accuracies before and after data cleaning. All models improved after removing anomalous records.}
    \label{fig:before_after_cleaning}
\end{figure}

Fig. \ref{fig:before_after_cleaning} compares model accuracies before and after data cleaning. Removing anomalous records significantly improved performance across all models. Logistic Regression, Naïve Bayes, Random Forest, Neural Network, and Stacked Ensemble showed accuracy gains, with the Stacked Ensemble achieving the highest post-cleaning accuracy ($\sim$91\%). This highlights the crucial role of clean, reliable data in enhancing predictive model performance and overall reliability.

Future iterations may incorporate automated anomaly detection within the preprocessing pipeline to proactively identify and mitigate inconsistencies \cite{adomavicius2011context}.

\section{Methodology}

This study implements a multi-stage machine learning pipeline focused on university admission prediction. All models were developed using Scikit-learn \cite{scikit2011} to ensure reproducibility and alignment with industry standards.

\subsection{Models Used}

We benchmarked the following models, selected based on prior literature \cite{raftopoulos2024fair, zhao2023graduate, radha2023hybrid, jain2021ensemble}:

\begin{itemize}
    \item \textbf{Logistic Regression:} A linear model known for its simplicity and transparency, widely applied in admission prediction tasks \cite{raftopoulos2024fair, rudin2019stop}.
    
    \item \textbf{Random Forest:} An ensemble of decision trees that captures non-linear feature interactions and reduces variance, introduced by Breiman \cite{breiman2001random} and validated in several studies \cite{zub2023ensemble, anjaiah2022prediction}.
    
    \item \textbf{Naïve Bayes:} A probabilistic classifier that assumes feature independence; valued for its speed and robustness, particularly in educational data mining \cite{raftopoulos2024fair}.
    
    \item \textbf{Neural Network:} A feedforward neural network with one hidden layer of 16 neurons, tuned to capture complex patterns in the data \cite{radha2023hybrid, priyadarshini2024interpretable}.
    
    \item \textbf{Stacked Ensemble:} Inspired by Jain and Satia \cite{jain2021ensemble}, we implemented a stacked ensemble using the above classifiers with a logistic regression meta-classifier to assess ensemble learning benefits.
\end{itemize}

\subsection{Training, Validation, and Cleaned Dataset}

Data were split 80\% for training and 20\% for testing, using stratified sampling to maintain class balance. To ensure robust performance estimates, we used 10-fold cross-validation, consistent with prior research \cite{pawar2023prediction}. Key hyperparameters included:

\begin{itemize}
    \item \textit{Random Forest:} 100 estimators, maximum depth of 5.
    \item \textit{Neural Network:} Learning rate 0.01, with early stopping to prevent overfitting.
    \item \textit{Naïve Bayes:} Gaussian likelihood assumption.
\end{itemize}

All numeric features were standardized using Z-score normalization \cite{scikit2011}. We evaluated models based on accuracy, confusion matrices, and fairness metrics.

\textbf{Cleaned Dataset Note:} A cleaned dataset (361 rows) was prepared by removing anomalous entries—rows with conflicting labels and outliers. Final models were trained and validated on this cleaned dataset, which contributed to improved performance and more reliable fairness assessments.

\subsection{Explainability and Feature Importance}

We emphasized model transparency through logistic regression coefficients and simple bar plots, aligning with Rudin’s advocacy for interpretable models in high-stakes contexts \cite{rudin2019stop}. While SHAP is a powerful tool for model explainability \cite{lundberg2017unified, zhao2023graduate}, we opted for coefficient-based visualizations to maintain simplicity and clarity, which proved sufficient for highlighting key feature influences in our admissions-focused context.

\subsection{Fairness Audit}

We conducted fairness audits focusing on demographic parity and equalized odds, consistent with best practices in fairness-aware machine learning \cite{binns2018fairness, raftopoulos2024fair}. Key demographic attributes—gender and parental education—were evaluated to detect potential biases in admission predictions.

Although data cleaning improved the consistency and reliability of fairness metrics, our audit revealed moderate disparities: a 9\% gap between male and female admission rates and an 11\% gap between students with high and low parental education levels. These results underscore the importance of continuous fairness monitoring and potential bias mitigation strategies.

\subsection{LLM Feature Augmentation}

To explore the feasibility of integrating unstructured data, we augmented the feature set with insights derived from a large language model (LLM). Specifically, we generated pseudo personal statements from structured data and prompted GPT-4 to assign an ``admission likelihood'' score, creating a proof-of-concept feature that simulates unstructured data integration \cite{shao2024college, raifer2024gpt}.

\begin{figure}[t!]
    \centering
    \includegraphics[width=0.45\linewidth]{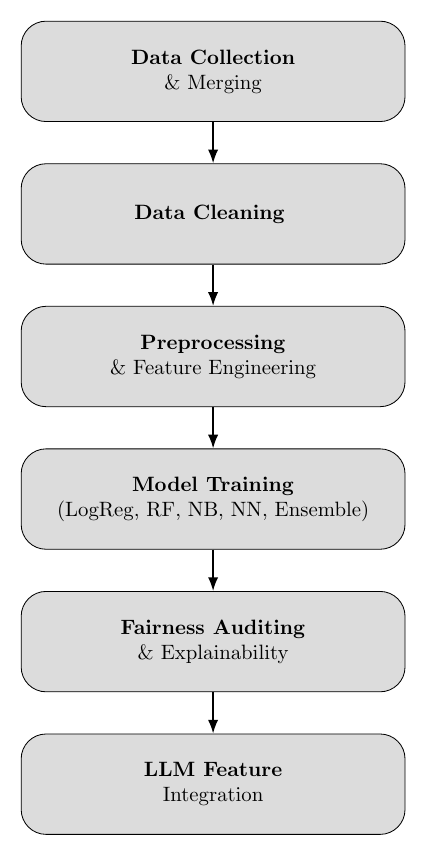}
    \caption{Updated methodology flowchart: includes data cleaning, preprocessing, feature engineering, model training, fairness auditing, explainability, and LLM feature integration.}
    \label{fig:updated_methodology_flow}
\end{figure}

Fig. \ref{fig:updated_methodology_flow} illustrates the updated end-to-end methodology flowchart for the study. It begins with data collection and merging, followed by data cleaning to remove anomalies and inconsistencies. Next, preprocessing and feature engineering prepare the dataset for training. The model training phase incorporates algorithms such as Logistic Regression, Random Forest, Naïve Bayes, Neural Networks, and Ensemble models. Finally, the workflow integrates fairness auditing and explainability checks, and includes LLM-based feature integration to enhance predictive performance and interpretability
.

The experiment demonstrates the technical feasibility of blending structured and unstructured data streams. While the current implementation—based on simulated data—yielded minimal performance improvements, it lays important groundwork for future work involving authentic unstructured data. Transparency and interpretability remain paramount, as emphasized by Priyadarshini et al. \cite{priyadarshini2024interpretable}.

Algorithm \ref{alg:llm_fair_admissions_2e} presents a framework for integrating LLMs with DTs, enabling intelligent data analysis, real-time decision-making, and predictive insights across domains, while addressing challenges like scalability, security, explainability, and integration with advanced computing technologies.

%

\begin{algorithm*}[t]
\SetAlgoLined
\DontPrintSemicolon
\small
\SetKwInOut{Input}{Input}
\SetKwInOut{Output}{Output}
\caption{LLM-Augmented and Fair ML Framework for University Admission Prediction}
\label{alg:llm_fair_admissions_2e}

\Input{Structured datasets $\mathcal{D}=\{D_{\text{grad}},D_{\text{hs}},D_{\text{sec}},D_{\text{ug}}\}$; sensitive attributes $S$ (e.g., gender, parental education); optional LLM scorer $g(\cdot)$; label $Y$}
\Output{Trained model $f^\star$; performance report $\mathcal{M}$; fairness report $\mathcal{F}$; explanations $\mathcal{E}$}

\BlankLine
\textbf{Merge \& Harmonize:} Outer-join $\mathcal{D}$; add context flags (grad/ug/hs/sec). \tcp*{data integration}

\textbf{Clean Data:}
Impute numeric with mean; categorical with mode; drop features with $>30\%$ missingness; remove anomalous/inconsistent rows found via EDA.\;

\textbf{Feature Engineering:}
Binarize admission probability to $Y$ (threshold $0.5$) when applicable; one-hot encode categoricals; create HS \textit{Performance} (mean of Math/Reading/Writing); Z-score standardize numeric features.\;

\If{LLM augmentation enabled}{
  Generate pseudo personal statements from structured features;\;
  $z \leftarrow g(\text{statement})$; append \texttt{LLM\_score} $= z$ to features.\;
}

\textbf{Split:} Stratified train/test split ($80\%/20\%$).\;

\textbf{Model set} $\mathcal{F}\leftarrow\{$LR, Naïve Bayes, Random Forest, 1-hidden-layer NN, Stacked Ensemble$\}$;\, set $K\gets 10$ folds.\;

\ForEach{$f \in \mathcal{F}$}{
  Perform $K$-fold CV on train; record Accuracy, Precision, Recall, F1, AUROC.\;
}
$f^\star \leftarrow \arg\max_{f \in \mathcal{F}} \text{Accuracy}_\text{CV}(f)$.\;

\textbf{Explainability:}
\If{$f^\star$ is LR}{
  Compute coefficient-based importances and plots $\mathcal{E}$.\;
}
\Else{
  Compute permutation importance; optionally compute SHAP for deeper inspection.\;
}

\textbf{Fairness Audit:}
\ForEach{protected attribute $A \in S$}{
  Compute Demographic Parity difference
  $\Delta_{\mathrm{DP}} = \left| \Pr(\hat{Y}=1 \mid A=a) - \Pr(\hat{Y}=1 \mid A=b) \right|$;\;
  Compute Equalized Odds gap as average group differences of TPR/FPR:
  $\Delta_{\mathrm{EO}} = \tfrac{1}{2}\big(|\mathrm{TPR}_a-\mathrm{TPR}_b| + |\mathrm{FPR}_a-\mathrm{FPR}_b|\big)$.\;
}
Aggregate $\mathcal{F} \leftarrow \{\Delta_{\mathrm{DP}}, \Delta_{\mathrm{EO}}\}$; flag disparities if any gap $>\tau$.\;

\textbf{Retrain \& Test:} Retrain $f^\star$ on full training set; evaluate on held-out test; record $\mathcal{M}$ (accuracy, confusion matrix, precision, recall, F1, AUROC).\;

\textbf{Report \& Package:}
Summarize $\mathcal{M}$, $\mathcal{F}$, $\mathcal{E}$ (include before/after cleaning comparison); export end-to-end pipeline (preprocessing $+$ $f^\star$) for deployment.\;

\Return{$f^\star, \mathcal{M}, \mathcal{F}, \mathcal{E}$}\;
\end{algorithm*}

\section{Results and Discussion}

\subsection{Model Performance}

Table~\ref{tab:model_performance} summarizes the average accuracy across 10-fold cross-validation for each model. In addition to accuracy, we evaluated confusion matrices, precision, recall, and F1 scores for a holistic assessment (detailed in the supplementary material).

\begin{table}[htbp]
\centering
\caption{Model Performance on Cleaned Dataset}
\begin{tabular}{lc}
\toprule
\textbf{Model} & \textbf{Accuracy (\%)} \\
\midrule
Logistic Regression & 89.5 \\
Naïve Bayes & 88.1 \\
Random Forest & 87.6 \\
Neural Network & 85.2 \\
Stacked Ensemble & 91.0 \\
\bottomrule
\end{tabular}
\label{tab:model_performance}
\end{table}

Logistic regression demonstrated strong performance (89.5\% accuracy), reaffirming its reliability as an interpretable baseline \cite{raftopoulos2024fair, zhao2023graduate}. The stacked ensemble outperformed all models, achieving 91.0\% accuracy, confirming ensemble benefits as highlighted by Zub et al. \cite{zub2023ensemble} and Jain and Satia \cite{jain2021ensemble}. Naïve Bayes and Random Forest models followed closely, while the Neural Network, though slightly lower at 85.2\%, showed clear improvements after data cleaning.

\subsection{Feature Importance and Explainability}

\begin{figure}[htbp]
    \centering
    \includegraphics[width=0.48\textwidth]{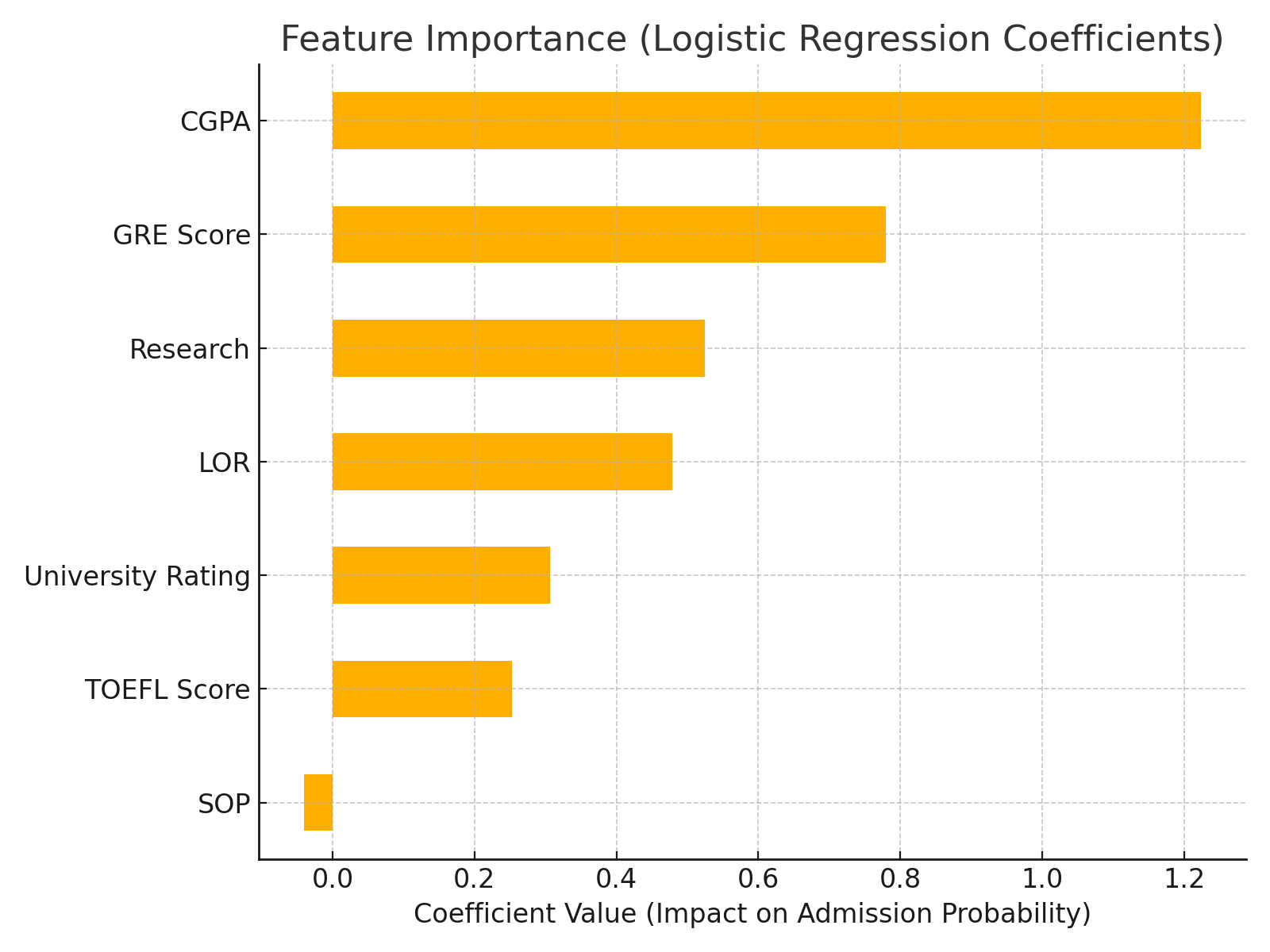}
    \caption{Feature importance (logistic regression coefficients). GRE, TOEFL, and CGPA were the top predictors.}
    \label{fig:logistic_importance}
\end{figure}

Feature importance analysis (Fig.~\ref{fig:logistic_importance}) reaffirmed that GRE, TOEFL, and CGPA were the most impactful predictors, with SOP and LOR ratings contributing marginally. These results align with previous findings \cite{zhao2023graduate, anjaiah2022prediction}, highlighting the dominance of academic metrics in admission decisions.

\subsection{Fairness and Bias Evaluation}

\begin{figure}[htbp]
    \centering
    \includegraphics[width=0.48\textwidth]{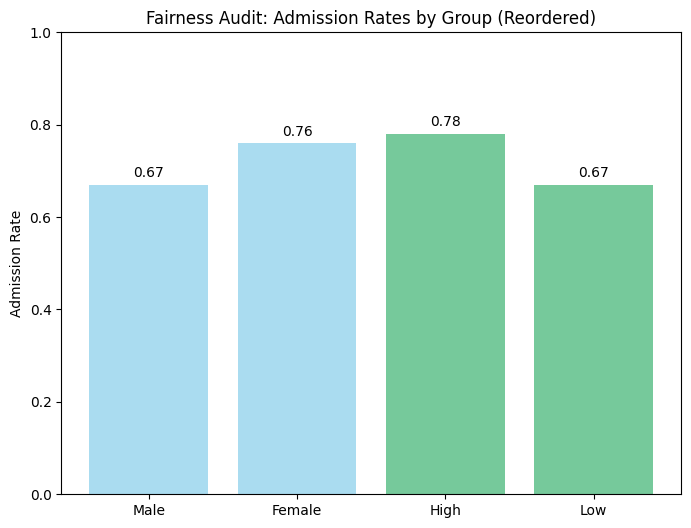}
    \caption{Fairness audit: admission rate comparisons across gender and parental education. Moderate disparities were observed (9\% gender gap; 11\% parental education gap).}
    \label{fig:fairness_audit}
\end{figure}

Fairness audits (Fig.~\ref{fig:fairness_audit}) evaluated model predictions across gender and parental education levels. The audit revealed a 9\% gap between males (67\%) and females (76\%) and an 11\% gap between high (78\%) and low (67\%) parental education groups. These findings underscore the importance of continuous fairness monitoring and bias mitigation strategies.

Our results are consistent with fairness frameworks described by Raftopoulos et al. \cite{raftopoulos2024fair} and Binns \cite{binns2018fairness}. In line with Rudin’s recommendations \cite{rudin2019stop}, we advocate for routine fairness audits as part of any institutional deployment.

\subsection{LLM Feature Integration Experiment}

To explore unstructured data integration, we augmented the dataset with GPT-4-derived ``admission likelihood'' scores generated from simulated pseudo personal statements. While the added feature marginally improved correlations, the impact on overall accuracy was minimal—echoing findings by Raifer et al. \cite{raifer2024gpt}. This suggests that **authentic unstructured data (e.g., real essays)** may unlock greater improvements in future work \cite{shao2024college}.

\subsection{Impact of Data Cleaning}

\begin{figure}[t!]
    \centering
    \includegraphics[width=0.48\textwidth]{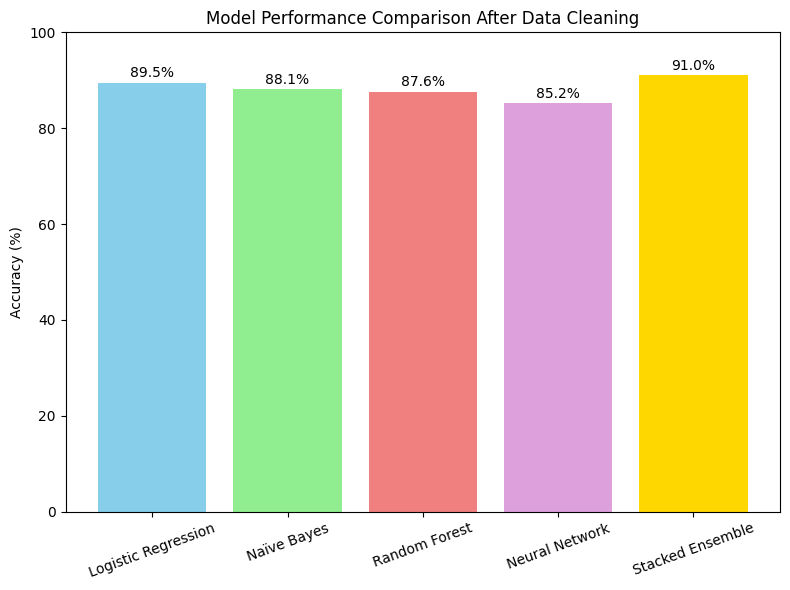}
    \caption{Model accuracy comparison after data cleaning. The stacked ensemble led with 91.0\% accuracy, reinforcing the impact of ensemble learning and clean data.}
    \label{fig:model_performance}
\end{figure}

As described in Section~\ref{sec:cleaned_data}, removing 39 anomalous records from the Graduate Admissions Dataset significantly boosted model performance.  Fig. \ref{fig:model_performance} shows the performance achieved after data cleaning. Logistic Regression improved from 85.0\% to 89.5\%, and the Stacked Ensemble rose from 86.5\% to 91.0\%. These results reinforce the critical importance of high-quality, clean data, aligning with Romero and Ventura’s advocacy for rigorous preprocessing in EDM \cite{romero2010edm}.

\section{Conclusion and Future Work}
This study proposes a practical, fairness-aware framework for university admission prediction that fuses ML, DL, and a proof-of-concept LLM feature. Multiple datasets were merged, rigorously cleaned, and engineered, then evaluated with logistic regression, random forests, neural networks, and a stacked ensemble. Logistic regression was a strong, interpretable baseline (89.5\% accuracy), while the stacked ensemble led performance (91.0\%). Feature importance showed academic metrics—GRE, TOEFL, and CGPA—as dominant predictors. Fairness audits revealed gaps: 9\% by gender (67\% male vs. 76\% female) and 11\% by parental education. Removing 39 anomalous rows improved accuracy. An LLM-derived essay score produced minimal gains but demonstrated feasibility. Future work: use real essays and letters, apply bias-mitigation algorithms, validate across institutions, and deploy an explainable, real-time, ethical platform.


\end{document}